\documentstyle[prb,aps]{revtex} 
\begin{document} 
\draft

\twocolumn[\hsize\textwidth\columnwidth\hsize\csname
@twocolumnfalse\endcsname

\title{
Calculation of the band gap energy of ionic crystals
}
\author{A. Aguado$^1$, A. Ayuela$^2$, J. M. L\'opez$^1$, J. A. Alonso$^1$, 
J. F. Rivas-Silva$^3$
and M. Berrondo$^4$}
\address{1. Departamento de F\'\i sica Te\'orica, Facultad de Ciencias,
Universidad de Valladolid, 47011 Valladolid, Spain. \\
2. Institut f\"ur Theoretische Physik, Technische Universit\"at Dresden,
01062 Dresden, Germany. \\
3. Instituto de F\'\i sica, Universidad Aut\'onoma de Puebla,
72750 Puebla, M\'exico. \\
4. Brigham Young University, Provo, UT84602, USA.}
\maketitle
\begin{abstract}
The band gap of alkali halides, alkaline-earth oxides, Al$_2$O$_3$ and
SiO$_2$ crystals has been calculated using the 
Perturbed-Ion model
supplemented with some assumptions for the
treatment of excited states. The gap is calculated in several ways:
as a difference between one-electron energy eigenvalues and as a
difference between the total energies of appropriate electronic states of the
crystal, both at the HF level and with inclusion of Coulomb correlation effects.
The results compare well with experimental band gap energies and with other
theoretical calculations, suggesting that the picture of bonding and excitation
given by the model can be useful in ionic
materials.

%
%
\end{abstract}



\vskip2pc]

\section{Introduction}

The increasing interest which is presently observed in luminescent
materials is due to their numerous technological applications, namely,
luminescent lighting and preparation of lamp phosphors, 
nuclear spectroscopy, laser science, or the construction of two-dimensional
detectors for use in medical screens and crystallography, to mention a 
few\cite{Bla95}.
The calculation of the properties of those materials requires a model for the
material and a computational model. The first one concerns the modelling
of the crystal of interest.
The computational model concerns the level of theory used in the calculations.
The calculation of the electronic properties of doped ionic crystals 
(or crystals containing vacancies) at an
{\em ab initio} level is still a challenge for the computational
methods normally used for crystalline solids or molecules.
The reason in the first case is the breaking of the translational symmetry
by the presence of
the defect. This enters in conflict with Bloch's theorem, which is the
basis of the solid state calculations. 
The use of molecular methods in real space requires 
a large number of atoms to be treated in a
selfconsistent way\cite{Ber95}. 
Those molecular
methods replace the whole crystal by a cluster, but the role of the
surface should not be overlooked even for large clusters.
An intermediate approach is provided by embedded cluster models, 
in which the impurity is surrounded by a
small fraction of the crystalline environment, and the rest of the crystal is
simply described by point charges. This approach has been already applied
to the problem
of luminescent impurities\cite{Ber95}.
Evidently, the description by point charges only gives an approximate
representation of the bulk.
In summary, a large number of atoms is needed for simulating an ionic crystal
containing an impurity, and even in such a case
the inherent problems of the cluster surface and the description of the
infinite bulk should be faced. Thus, a model that gives a 
better description of the
impurity-lattice and cluster-lattice interactions is desirable.
The ideas of the Theory of Electronic Separability
(TES) developed by Huzinaga and coworkers\cite{Huz71,Huz73,Lua87}
should be useful in this context. The TES supplies a
natural framework to develop accurate schemes for dealing with the
cluster-lattice interaction.
The perturbed ion (PI) model\cite{Lua90}, based on the TES,  
was developed for the study of ionic crystals or,
more in general, for crystals formed by atoms with closed shells,
 and has been later extended to free clusters of the same
materials\cite{Ayu93,Agu97}. 

We plan to apply the PI model to study luminescent centers in ionic crystals.
As a first step, we use the PI model in this paper to calculate
the band gap of pure alkali halide, alkaline-earth oxide, Al$_2$O$_3$ and
SiO$_2$ crystals. All these materials are important in the field of
luminescence and also in other fields ranging from catalysis to
magneto-optical devices. 
The
calculation of the energy band gap of these materials 
is important for several reasons: first of all, 
the impurity levels of the doped crystals are located
in the band gap of the pure crystal.
Furthermore, the gap is a very important quantity in the
first step of the scintillation process, namely, absorption of
radiation leading to formation of electron-hole pairs. 
This step influences the global 
efficiency of the scintillator. Last but not the least, 
it is also important
from the theoretical point of view, as it gives information on the quality of
the model. 
A large body of work exists on 
the calculation of the band gaps in these materials using the traditional
methods of band 
theory\cite{Cla68,Kun68a,Fon68,Kun68b,Fow69,Pag70,Kun70,Ove71,Lip71,Per72,Poo75a,Jou75,Kun82,Hea83,Pan91}, 
and we compare our results to a 
representative set of band calculations. 
Although we obtain a better global agreement with experimental gaps than
many {\em ab initio} band structure calculations,
our intention here is not to
compete with those well established solid state methods. Instead, we only 
intend to show that the PI model affords an accurate description of the band
gap within the framework of a cluster-like  approach, which is considered a
convenient approach for the study of doped crystals.

The structure of the paper is as follows: In section II we present the PI 
model
for ionic crystals, showing how several problems concerning the cluster
approximation are circunvented. 
Since the PI model is originally formulated in a
Hartree-Fock(HF) framework, we also discuss 
the introduction of Coulomb correlation.
Section III describes the calculation of the gap. Results are presented for
the above mentioned crystals 
and the trends obtained are discussed in comparison
with experiment and other calculations.
Finally, section IV summarizes our conclusions.

\section{The Perturbed Ion model}

\subsection{Theory}

The PI model has been developed for 
systems (pure crystals, crystals with defects, or
finite clusters) formed by weakly correlated one-center electronic groups,
the prototypical systems being ionic crystals, 
like the alkali halides, formed
by closed-shell ions.
Therefore, according to the 
theory of electronic separability\cite{Huz71,Huz73,Lua87}, the wavefunction
of the system
can be expressed as an antisymmetrized product of the local wavefunctions
describing each group. 
If these local wavefunctions satisfy 
strong-orthogonality conditions\cite{Lyk56,Par56}, the
total energy is the sum of intragroup, or net energies, and intergroup,
or interaction energies: 

\begin{equation}
E_{\em system}= \sum_{R} E_{net}^{R} + 
\frac{1}{2} \sum_{R} \sum_{S(\not= R)} E_{int}^{RS}
\end{equation}
where the R and S sums run through local groups.
All the contributions to $E_{\em system}$ due to a group
A can be collected in an effective energy defined as

\begin{equation}
E_{eff}^A=E_{net}^{A}+ \sum_{S(\not= A)} E_{int}^{AS}=E_{net}^A+E_{int}^A,
\end{equation}
but it
should be noted that $E_{\em system}$ is not a simple sum of the effective
energies since the interaction energies are then counted twice. So the 
additive energy of a group $A$ is defined as

\begin{equation}
E_{add}^A=E_{net}^A+\frac{1}{2}\sum_{S(\not= A)}E_{int}^{AS}=E_{net}^A+
\frac{1}{2}E_{int}^A
\end{equation}
and then the total energy of a system formed by {\em a} groups of type {\em A},
{\em b} groups of type {\em B}, etc, can be written 

\begin{equation}
E_{\em system}(A_aB_b...)= aE_{add}^A + bE_{add}^B + ...
\end{equation}
In practice, for ionic crystals the local 
groups will be identified with the closed-shell ions.

Group wavefunctions are
obtained by minimizing their effective energies if strong orthogonality
conditions are satisfied among the group wavefunctions. This restricted
variational procedure can be successive 
and iteratively applied to all the different groups, in
order to determine fully consistent group wavefunctions and 
the best system wavefunction compatible with the initial
assumption of separability. When the wavefunction of a particular group is
being determined, the wavefunctions of all the other groups are
considered frozen.

The effective energies
can be expressed as  expectation values of appropriate effective Hamiltonians.
For group {\em A}
\begin{equation}
E_{eff}^A=<\psi_A \mid H_{eff}^A\mid \psi_A> 
- \sum_{S(\not= A)} Z^A V_{H}^S (R_{AS}).
\end{equation}
The group-wavefunction $\psi_A$ is a Slater determinant. 
The first term in eq.(5),
namely $<\psi_A \mid H_{eff}^A\mid \psi_A>$, 
collects the energy associated to the
electronic cloud of group {\em A}, and the second term gives the interaction
between the nucleus $A$, of charge $Z^A$, and the Hartree potential of
groups $S(\not=A)$. $R_{AS}= \mid \vec R_A - \vec R_S\mid$ indicates the
distance between the two nuclei. The effective Hamiltonian $H_{eff}^A$ can be
written as a sum of several terms
\begin{eqnarray}
H_{eff}^A=\sum_{i=1}^{N_A}T(i) - 
\sum_{i=1}^{N_A}Z^A r_{iA}^{-1} + \nonumber \\
\sum_{1\le j< i\le N_A}r_{ij}^{-1} +
\sum_{i=1}^{N_A} \sum_{S(\not=A)}[V_{eff}^S (i) + P^S (i)],
\end{eqnarray}
with $r_{iA} = \mid \vec r_i - \vec R_A\mid$ and 
$r_{ij} = \mid \vec r_i - \vec r_j\mid$. The i and j-summations run
over the $N_A$ electrons of group {\em A}. The first three terms in the
Hamiltonian represent the electronic kinetic energy, 
the interaction between the electrons
and the nucleus and the interelectronic repulsion (those three terms refer
exclusively to group {\em A}). 
On the other hand, the terms containing the 
S-summation account for the interaction with the frozen groups {\em S}. This
interaction separates itself in two parts. First,
$V_{eff}^S(i)$ represents the effective potential energy
of an electron (of group A) in the mean field of the group $S$:
\begin{equation}
V_{eff}^S (i) = - Z^S r_{iS}^{-1} + V_C^S (i) + V_X^S (i) = 
V_{\em H}^S(i) + V_X^S(i),
\end{equation}
where the three terms in $V_{eff}^S$ are the electron-nucleus,
the classical electron-electron and the
exchange parts of the potential energy (the sum of the first two terms
is the Hartree potential $V_H^S$).
Second, the strong orthogonality between the orbitals of 
the active group A and those of the other groups {\em S}
is 
enforced  in H$_{eff}$ by means of the projection operator P$^S(i)$.
For systems formed by closed shell ions, this operator 
takes the form\cite{Huz87} 
\begin{equation}
P^S(i) = \sum_{g \in S} \mid \phi_g^S> (-2 \epsilon_g^S ) <\phi_g^S \mid,
\end{equation}
where g runs over all occupied one-electron orbitals 
$\phi_g$ (with orbital energies
$\epsilon_g^S$) of group {\em S}.
 
We now deal with the explicit form of $V_{eff}^S$\cite{Lua90}. 
For the  closed-shell ions  considered here 
the Hartree part of this potential is given by
\begin{equation}
V_{\em H}^S(\vec r_1)= -Z^Sr_{1S}^{-1} + 
\int \rho^S (\vec r_2) r_{12}^{-1} d\tau_2
\end{equation}
where $\rho^S(\vec r)$ is the electron density of the
ion. It is physically and computationally 
convenient to separate this electrostatic 
potential into classical and nonclassical (nc) terms such that:
\begin{equation}
V_{\em H}^S(\vec r_{1S})= -q^Sr_{1S}^{-1} + V_{nc}^S(\vec r_{1S})
\end{equation}
where $q^S$ is the net charge of the ion, and $V_{nc}^S$ represents the
deviation of $V_{\em H}^S$ from a point 
charge potential due to the finite extension of the
electronic density of the ion. This operator can be effectively computed using
the Silverstone expansion of a function in a displaced center\cite{Sil77,Map91}.

The exchange operator $V_X^S$ can be written 
as the nondiagonal
spectral resolution\cite{Huz87}
\begin{equation}
V_X^S(i)=-\sum_l \sum_{m=-l}^l \sum_{a,b} \mid alm,S> A(l,ab,S) < blm,S\mid.
\end{equation}
Here $\mid alm,S>$ are products 
of spherical harmonics $Y_l^m$ and primitive radial
functions for the S ion, $a$ and $b$ 
run over the Slater-Type Orbitals (STO) of 
$l$ symmetry, and $A(l,ab,S)$ are the elements of the matrix
\begin{equation}
A={\cal S}^{-1} {\cal K} {\cal S}^{-1}
\end{equation} 
where ${\cal S}$ and
${\cal K}$ are the
overlap and the exchange matrices for the S ion in the \{$\mid alm,S>$\}
basis.
Full details can be found in
the original paper by Lua\~na and Pueyo\cite{Lua90}.

As indicated above, when the effective Hamiltonian 
of an active group {\em A} is
being diagonalized, all the other groups {\em S} are considered frozen.
Selfconsistency is achieved following an iterative
scheme.  
At each iteration all the inequivalent ions are successively treated as
active groups, and the orbital wavefunctions and eigenvalues of the frozen
groups are taken from the previous iteration.  The iterations are continued
until convergence is achieved.

Some features of the PI model are worth mentioning here.
In the cluster approaches the
solid is partitioned into a ``cluster region'' and ``the rest of the
crystal''. The physical and mathematical description of both entities should
be accurate enough, as well as their interrelations. 
Frecuently, the cluster has been
solved using precise quantum-mechanical methods while the environment has been
simulated by using point charges\cite{Bar83}, but more accurate descriptions
of the environment surrounding the active 
cluster are necessary in order to account
properly for cluster-lattice interactions\cite{Kun88}.
Besides, the 
election of the cluster size is also a delicate problem, because surface
effects at the cluster boundary may affect the results. 
In the PI model, the ``cluster'' is reduced to its minimum
size, a single ion, and cluster-lattice interactions are described 
selfconsistently in the
framework of the TES\cite{Huz71,Huz73}.
The cluster approximation can be rigourously formulated within the
TES, as cluster-lattice orthogonality is a fundamental requirement of that
theory.
Another feature of the PI model is that it does not invoke the
LCAO approximation.
The one-center character of the model
leads to a large computational saving with respect to any multi-center
cluster approach.
The PI model just
described has been formulated at the Hartree-Fock level. 
Now we turn to the introduction of
Coulomb correlation.

\subsection {Introduction of correlation in the PI model. 
The Coulomb hole model}

The seminal idea of separability is to find electronic groups such that
intergroup correlations play a minor role\cite{McW94}. Intraatomic
correlation, on the other hand, 
contributes significantly to the band gap energy
of ionic crystals and has to be included.  To take into account
intraatomic
correlation we have used the Coulomb-Hartree-Fock (CHF) model proposed
by Clementi\cite{Cle65,Cha89}.
In the CHF model, the coulomb-repulsion integrals are
modified by introducing a 
spherical hole around each electron in which the other
electrons do not penetrate.  The radius of this coulomb hole for a particular
integral depends on the overlaps between the functions involved,
aside from other factors. Two
parameters can scale the hole, and their values were chosen to match the
empirical correlation energies of He and Ne.  Details of the CHF model can be
consulted in Clementi's papers\cite{Cle65,Cha89}.

From the point of view of the PI calculations, the CHF model is convenient for
several reasons: (a) it is computationally simple; (b) it can be implemented
easily
in the Roothaan-Bagus SCF (Self Consistent Field)
formalism\cite{Roo63}; (c) it reproduces within a few parts per cent the 
empirical correlation energy for the 
ground state of free atoms\cite{Cle65,Cha89};
(d) it is
size consistent; and (e) the CHF correlation energy depends on the radial
density of the ions, thus incorporating correlation effects upon the formation
of the solid.

There are two ways to implement the CHF model. Without going into more detail,
we have used the unrelaxed CHF approximation, in which the PI wavefunctions
calculated at the HF level are kept fixed and the correlation energy
is simply added to the HF energies.
Since the coulomb repulsion integrals are reduced 
by the effect of the hole, the
corrected energies of the ions are smaller than their corresponding 
original PI
values. The good performance of the CHF model in accounting for intraatomic
correlation can be appreciated in the reviews of refs.\cite{Pue92,Lua92a}.

\subsection{Basis set}

The localized atomic-like orbitals used to describe each ion in the
crystal are expanded in a large
basis set of STO's\cite{Cle74,Mcl81}
because of their superior performance (their precision
being near Numerical-Hartree-Fock).
An extension of the basis was required for the description of
bromides. From an analysis of preliminary calculations we have reasons to 
think that the basis set used for $Br^-$ may not
be the most appropriate to 
describe this anion in the crystalline environment.
Here we have decided to enlarge the basis
set for $Br^-$ by adding one
5p-polarization function. The exponent
of the 5p-STO was adjusted by minimizing the total energy
of the corresponding crystal. We
have found that upon inclusion of the 5p-STO, 
the 4p-orbital of $Br^-$ experiences a
small contraction with respect to the 4p-orbital in the absence of the
polarization function. This very small effect is responsible for the
improvement of the crystal energy.  
Within the TES,
the portion of the total energy
represented by the projection energy is sensitive to the quality of the
basis set in the tail region.
Further discussion on the influence of basis sets on the results
is provided in section III. 

\section {Calculation of energy gaps}

Following the work of Poole {\em et al.}\cite{Poo75a}, 
the gap is rigorously defined
as the difference between the 
threshold energy $E_t$ and the electronic affinity
$\chi$ of the cristal:
\begin{equation}
\Delta E_{gap} = E_t - \chi.
\end{equation}
The threshold energy is the energy needed to remove an electron from the top
of the valence band (VB), whereas the electronic affinity is defined as the 
energy of an
electron at the bottom of the conduction band (CB), referred to the vacuum
level. It is important to be aware of this definition, because if the hole in
the VB and the electron in the CB were allowed to interact (electron-hole
pair), strictly we would
be describing instead an excitonic level.

The problem for a cluster model when trying to obtain band gap energies using
eq.(13) is the calculation of the electronic affinity $\chi$, as it is
necessary to deal with an electron in a delocalized state
at the bottom of the conduction band\cite{Bag94}.
The PI model
is not an exception in this respect. 
Due to the one-center character of the model, we have to center
the electron wave function on a lattice site. Besides, the strong ion-lattice
orthogonality required by the TES 
would force the electron to be localized on a 
given ion, leading to an incorrect representation of
a conduction band state.
We can, however, give an 
approximate description of this state by relaxing the orthogonality
requirements (see below). For ionic materials
the experimental values of
$\chi$ are small 
(only a few tenths of eV\cite{Poo75a,Poo75}) compared to $E_t$ and our
approximation
gives values for $\chi$ within the correct order of magnitude.
Besides, typical errors in measured gaps are $\sim$ 0.5
eV.

We have calculated the band gap 
energy in two different ways. 
The first one identifies the gap with
the difference between the energy eigenvalues corresponding to the lowest
unoccupied molecular orbital
(LUMO) and the highest occupied molecular orbital (HOMO)
obtained in the PI model at the HF level,
that is, we approximate the threshold energy $E_t$ by the HOMO and the
electronic affinity $\chi$ by the LUMO (with opposite signs):
\begin{equation}
\Delta E_{gap} = \epsilon(LUMO) - \epsilon(HOMO).
\end{equation}
This energy
difference overestimates the gap, as is the case of typical band theory 
calculations\cite{Kun82}. The second 
way is a $\Delta$SCF
calculation.

\subsection {LUMO - HOMO difference.} 

The orbital energy of the HOMO level is a quantity readily obtained in the PI 
model. For a given alkali halide
crystal it corresponds to the eigenvalue of the outermost
occupied p-orbital of the anion. To obtain the LUMO we can simulate
a neutral alkali atom, $A^{0}$, as an impurity in the 
field created by the
pure crystal ($A^{0}:A^+X^-$). Then, the LUMO is identified with the
outermost occupied orbital of the neutral alkali impurity.
As stated above,
when the calculation is performed within the strict framework of the PI model,
the localized basis and the condition of strong 
orthogonalization 
between the
orbitals in neighboring sites would
lead to an unphysical localization of this electron
on the alkali site.
A better description of the LUMO is achieved by
freezing the crystal around the $A^{0}$ impurity (that is the wave functions
of the surrounding ions) and
removing the pieces $V_X^S (i)$ and $P^S(i)$ out 
of the effective Hamiltonian for the impurity.
It has to be stressed
that our calculation describes the LUMO as an occupied orbital centered on
an alkali atom. To allow for the delocalization of this orbital over a
substantial region of the crystal, the basis set used for 
the impurity (taken from Bunge and Barrientos\cite{Bun93} for Li, Na, K, Rb,
and from McLean\cite{Mcl81} for Cs) has been
enlarged by adding
some diffuse s-type STO functions, assuming that
their exponents form a geometric progression\cite{Raf73}. 
For the ratio of the geometric
progression we took
the ratio between the exponents of the two outermost s-orbitals of the original
basis set, and we enlarged successively 
the basis set with one, two and three STO's.
The calculated LUMO eigenvalue
converges fast as the basis set is enlarged. In fact, 
by adding a fourth diffuse function, the
LUMO energy changed by $\sim 10^{-3}$ eV only.
With the basis chosen,
the electron in the LUMO orbital of alkali halides spreads over 
a range of about 32 coordination
cells
from its center.
By removing the pieces $V_X^S(i)$ and $P^S(i)$ from the effective hamiltonian
for the impurity, our calculation of the LUMO becomes rather similar to the 
method used in some embedded cluster models, in which the metal impurity,
with an extra electron added, is surrounded by a set of point charges chosen
and optimized to represent the Madelung potential\cite{Cas93}. The difference
between the two procedures is that in our calculation we employ the complete
charge distribution of the crystal and not ionic point charges.

The calculation is similar for the oxide crystals. $\epsilon_{HOMO}$
is the eigenvalue of the highest occupied anionic orbital in the pure
crystal, and $\epsilon_{LUMO}$ is obtained by simulating a metal (or silicon)
cation with an electron added in its external s shell as an impurity in the
pure crystal. Unlike the case of alkali halides, in the oxides the metal
(or Si) cation remains charged (Be$^+$, Mg$^+$, Ca$^+$, Sr$^+$, Ba$^+$,
Al$^{2+}$, Si$^{3+}$) after addition of that electron. The 
Clementi-Roetti\cite{Cle74} basis for the cations, enlarged by some diffuse
functions, were used to simulate the delocalization of the LUMO orbital.

The magnitude of the LUMO eigenvalue is always a small quantity (the
largest value is 0.4 eV for
$Si^{3+}:SiO_2$).
Our results concerning the
affinity level agree with those obtained for $MgO$
by Bagus {\em et al.}\cite{Bag94} in the sense that this level bears no
simple connection with cationic s-orbitals. In all cases it was irrelevant to
center the electron wave function on a cationic or an anionic site, 
the only important issue was
to allow for the
necessary delocalization of the electron.
The number of point-charge shells included in our calculation to converge
the Madelung term is about 190, and the picture that arises
is that the electron becomes as delocalized as possible, the same
conclusion reached by Bagus {\em et al}\cite{Bag94}.
 
Results for the calculated band gaps of twenty alkali halide
crystals are 
compared in Table I with experiment\cite{Kun82,Poo75,Hay85} and with other 
calculations. 
In a similar way, Table II shows the results for seven oxide crystals, also
compared with experiment\cite{Web86,Str73,Ara68,Him86} and other theoretical
results\cite{Pan91,Lor93}.
Typical errors
in the experimental gaps are 0.2-0.5 eV\cite{Kun82}. 
Our calculations were 
performed using
the experimental geometries
and lattice constants of the perfect crystals\cite{Wyc63,Kit76}.
The 
calculated $(LUMO - HOMO)$
gaps display the experimental trends: a decrease of the band gap by  
moving down the periodic table along the cation column when the anion is
fixed (the experimental gaps for Br and I-crystals show some exceptions
to this trend),
and also a decrease by moving down the anion column when the cation
is fixed. For the alkali halides, we obtain a general 
overestimation of the gap, with errors
which increase with the size of the ions involved. The same trends are
observed in a calculation by Berrondo and Rivas-Silva\cite{Ber96} using a
model of a cluster embedded in a background of point-like ions.
Berrondo's calculation was performed 
at the HF level and
the gap, obtained as the difference of LUMO and HOMO orbital energies, is 
given in Table I. 
The error in the present approximate calculation is, however, about
1 eV smaller than in Berrondo's calculation.
On the other hand,
for oxide crystals, the $(LUMO - HOMO)$ calculation yields excellent
agreement with experiment. At this level of theory, correlation and orbital
relaxation effects are not yet taken into account, so this exceptional
agreement has to be considered as resulting from a subtle cancellation of
errors.

\subsection {$\Delta$SCF calculation.} 

The energy gap of an N-electron system is rigorously defined within the
$\Delta$SCF method as
\begin{equation}
\Delta E_{gap} = (E_{N-1} - E_N) - (E_N - E_{N+1}),
\end{equation}
where $E_N$ is the ground state energy\cite{God88}. This is, evidently,
equivalent to eqn. (13). For the alkali halide crystals we can write
\begin{eqnarray}
E_t = E_{N-1} - E_N = \nonumber \\ E_{crystal}(X^0:A^+X^-) - E_{crystal}(A^+X^-),
\end{eqnarray}
\begin{eqnarray}
\chi = E_N - E_{N+1} = \nonumber \\ E_{crystal}(A^+X^-) - E_{crystal}(A^0:A^+X^-),
\end{eqnarray}
associated to the two independent processes of removing an electron from a
halogen anion and of placing an electron in a state 
at the bottom of the conduction band, 
respectively. $E_{crystal}(X^0:A^+X^-)$ represents the energy of the crystal
with a single neutral halogen impurity and $E_{crystal}(A^0:A^+X^-)$ that of
a crystal with a ``neutralized'' alkali ion (although the
neutralizing electron is in a fully delocalized state). 
Finally $E_{crystal}(A^+X^-)$ is the energy of the perfect crystal.
The geometry of these ``doped''
crystals is that of the perfect crystal, 
according to the Franck-Condon principle.

The electronic orbitals of the ions surrounding the neutral
impurities (in fact, of all ions in the crystal) 
were kept frozen in the same electronic states as in the pure
crystal. This is not strictly necessary for the case $X^0:A^+X^-$ but we have
found from test calculations that the electronic relaxation of the 
cations surrounding the neutral halogen atom is 
indeed negligible. For the case of
$A^0:A^+X^-$, freezing of the electronic clouds of the surrounding shells of
ions is convenient in view of the treatment of the impurity (see below).
With these assumptions the energies $E_t$ and $\chi$ reduce to a
difference of effective energies\cite{Map92} 
\begin{equation}
E_t = E_{eff}(X^0:A^+X^-) - E_{eff}(X^-:A^+X^-) 
\end{equation}
\begin{equation}
\chi = E_{eff}(A^+:A^+X^-) -  E_{eff}(A^0:A^+X^-) 
\end{equation}
where the effective energies of $X^-$ and $A^+$ refer, evidently, to
ions of the perfect crystal.
Again, as in the discussion of the 
HOMO-LUMO gap in section III.A above, we have
enlarged the basis set to be able 
to describe an electronic state formally centered on
an alkali site although 
delocalized over a region corresponding to many shells
of neighbours.  
Applying the same argument to alkaline-earth oxides, the equivalent of
equations (18) and (19) are
\begin{equation}
E_t = E_{eff}(O^-:A^{2+}O^{2-}) - E_{eff}(O^{2-}:A^{2+}O^{2-}) 
\end{equation}
\begin{equation}
\chi = E_{eff}(A^{2+}:A^{2+}O^{2-}) - E_{eff}(A^+:A^{2+}O^{2-}),
\end{equation}
and related expressions can be easily written for the $SiO_2$ and 
$Al_2O_3$ crystals.

The main contribution to
$\Delta E_{gap}$ comes from $E_t$, and $\chi$ only provides a
small correction. The largest value of $\chi$, which is $\sim 0.5$ eV 
is found for $SiO_2$.
This magnitude is consistent with the
energy difference between the vacuum
level and the lowest conduction band state measured by photoemission
spectroscopy\cite{Poo75}. That energy difference is generally 
smaller than 1 eV, and
in most cases much lower.

The gaps obtained by this $\Delta SCF$ 
calculation are given in tables I and II.
Two sets of results are included: 
the first one corresponds to calculations at
the Hartree-Fock level, and the second includes correlation via the CHF
model. The gaps obtained for alkali halides are smaller than
those calculated by substracting LUMO and HOMO eigenvalues, 
a fact that improves the
agreement with experiment, in particular for cases involving
heavy halogens. The improvement arises from
orbital relaxation in 
response to the removal of one electron from the anion. The
relaxation of the orbitals of 
the halogen atom is included now in the calculation
of $E_t$, but not in the calculation of the HOMO level in section III.A.
This effect of a 
higher sensitivity of the anions (as compared to cations) to
orbital relaxation effects was also found in our previous works on
clusters\cite{Ayu93,Agu97}.
For oxides, improvement with respect to the $LUMO-HOMO$
difference of eqn. (14) is only found after including correlation.

The electron density of the cations is very localized and
practically does not change when the 
delocalized electron is centered on those
cations. Furthermore, the delocalized electron does not interact
with itself. This means that correlation gives a negligible contribution
to $\chi$, so one can concentrate on the contribution of correlation to
$E_t$. The difference between $\Delta SCF(HF)$ and $\Delta SCF(CHF)$
in Tables I and II comes from this source.
We can appreciate in Table I that the correlation correction 
improves the gap
for most alkali halides,
with KF, KCl, RbCl and CsCl as the only exceptions. Furthermore, it is
worth noting that this correction changes the magnitude of 
the gaps in the correct direction, namely,
increasing the gap energy for fluorides, which have HF gaps
below the experimental values, 
and decreasing the gap for the rest of the crystals,
which show (except KCl, RbCl and CsCl) HF gaps larger than the experimental 
ones. In summary, the global 
trends remain the same as in the HF calculations, but the
quantitative agreement improves with the inclusion of correlation effects.
The same can be said of the oxide crystals, where inclusion of correlation
reinstates the initial agreement found in the $(LUMO - HOMO)$ calculation.
These results lead us to conclude that correlation corrections are
important for a quantitative comparison with experiment, 
although the general trends are
reproduced at the HF level.

The band gaps obtained by Kunz\cite{Kun82} for twelve alkali halide crystals 
and by Pandey {\em et al.}\cite{Pan91} for three oxides from band structure
calculations 
are given for comparison in Tables I and II. Energy band calculations at
the HF level overestimate the band gap with errors of a few eV
(see ref.\cite{Kun82});
notice that this is a LUMO-HOMO-type of estimation.
Kunz\cite{Kun82,Pan91} then introduced coulomb correlation 
by updating a method described by
Pantelides {\em et al.}\cite{Pan74}. 
Correlation effects lowered the magnitude of the
gap, bringing the results (included in Tables I and II)
in good agreement with experiment.
Comparing with the results obtained by Kunz,
we can see that our CHF results are slightly better, except
for a few crystals (NaF, KF, RbBr), and that
Kunz calculations are less successful in
reproducing the general trends. 
We have just mentioned that HF band structure calculations
overestimate the gap by a sizable amount. On the other hand our $\Delta$SCF
calculations at the HF level do not seem to be much affected by this problem.
We ascribe this to electronic relaxation effects.
 
In Table I, we also include the
results of Berrondo and Rivas-Silva\cite{Ber96},
obtained as the difference between the
total energies of the lowest electronic 
excited state and the ground state of a cluster simulating a
piece of the ionic crystal, embedded in a system of point charges.  
These calculations were performed at the HF level and, evidently, include
electronic relaxation, so in Table I the corresponding results have been
classified under the $\Delta$SCF category. Berrondo's calculations include in
addition relativistic effects. A comparison with the present 
$\Delta$SCF(HF) results
shows rather similar gaps. In the same paper
Berrondo and Rivas-Silva have obtained the gap by identifying this one with
the excitation energy obtained in a Configuration Interaction (CI) calculation,
again within a cluster model, in this case a bare cluster without
surrounding point charges. As a consequence of the introduction of correlation, the
CI calculations
give an improvement of the calculated gaps for chlorides, bromides and
iodides. 
In the case of fluorides, where some of the 
$\Delta$SCF(HF) gaps obtained by Berrondo are below the
experimental values, the CI calculation leads to 
an increase of the gaps. So 
turning from $\Delta$SCF to CI seems to work in the right direction,
although for fluorides this results in a 
worsening of the magnitude of the gap. 
The errors in the CI gaps are not larger than 1 eV. This is also the case
for our $\Delta$SCF calculations including correlation, except for NaF, KCl
and RbCl, for which the error is 1.1 eV, and for CsCl and CsBr, where the
errors are 1.6 eV and 1.9 eV respectively. The computer time 
required by our calculations is, however, much
smaller.

Included in Table II are the theoretical results obtained by Lorda
{\em et al}\cite{Lor93} for alkaline-earth oxides. 
They used a model of a linear triatomic M-O-M cluster
(M= metal, O= oxygen) embedded in a Madelung field representing the
crystal. The {\em ab initio} wavefunction was written as a linear combination
of Slater determinants, each one corresponding to a resonating valence bond
structure. In that way a valence bond wave function is obtained which is
explicitely correlated from the outset, in the sense that it is not an 
eigenfunction of a 
hamiltonian which can be written as a sum of monoelectronic operators (for
details, consult chapter 7 of ref.\cite{McW94}). 
In order to include external
correlation effects they used standard second-order perturbation techniques.
This external correlation is defined as the energy 
contribution of those Slater
determinants not explicitely included in the valence bond model space.
The gap was then calculated as an
energy difference between appropriate resonating valence bond components.
Our results are also in better
agreement with experiment in this case.
Actually, the first
excitation energies calculated by Lorda {\em et al.}
should not be identified with the band gap
energies, but with localized excitations. This
point was emphasized when we gave 
the definition of the band gap energy above.

Density functional theory (DFT)\cite{Mar83} has also been used to calculate band gaps in
ionic crystals. The DFT one-particle eigenvalues have no formal
justification as quasi-particle energies although in practice these eigenvalues
have been used to discuss the spectra of solids. The DFT band gaps (usually
calculated employing the local density approximation for exchange and
correlation effects) give values typically 30-50 \% less than the gaps
observed in the optical spectra. A rigorous formulation requires the
calculation of the quasi-particle band structure. This has been done by Hybertsen
and Louie\cite{Hyb86} by using the so-called GW approximation to the
self-energy\cite{Hed65}. These authors have obtained a gap of 9.1 eV for LiCl.

\section{Conclusions.}

A preliminary requirement for the treatment of
luminescent impurities in ionic crystals
is an accurate description of the
band gap. Consequently,
we have carried out a series of approximate calculations of the band gap for
twenty alkali halide and seven oxide crystals. 
Two different prescriptions have been used: the
first one is a difference between the energies of the LUMO and HOMO orbitals
of the crystal, and the second is a $\Delta$SCF calculation based on
subtracting total energies.
For this purpose the Perturbed Ion
(PI) model has been used, supplemented with some additional assumptions to
deal with the state corresponding to one electron at the bottom of the
conduction band. Correlation effects have been taken into account by a model
introduced by Clementi.

Overall we have obtained a good description of the energy gaps with a very
modest computational effort. The gaps show a reasonable agreement with
experiment, 
and also compare well (in many cases favorably) with previous
theoretical works. The results support the validity of the PI model and the 
additional assumptions employed to simulate the physical situation. Further
improvement of some particular points (as the modelling of the delocalized
electron at the bottom of the conduction band, the treatment of electronic
correlation, and the introduction of relativistic effects for heavy atoms) are
expected to improve further the results.

$\;$

$\;$

{\bf Acknowledgments}:
This work has been supported by DGICYT(Grant PB95-0720-C02-01).
A. Aguado is greatful to Universidad de Valladolid, Caja Salamanca y
Soria, and Junta de Castilla y Le\'on for Postgraduate
fellowships. A. Ayuela thanks the hospitality and support from Brigham Young
University during a visit at the starting of this work.

\pagebreak

{\bf Captions of tables}

$\;$

{\bf Table I}
Band gaps of alkali halide crystals
calculated by different theoretical methods, compared to experimental
values\cite{Kun82,Poo75,Hay85}. 
$\Delta \epsilon_{HF}$(LUMO - HOMO) is a difference of
one-particle orbital energy eigenvalues. $\Delta$SCF refers to a difference
between the energies of the crystal in appropriate electronic states (see
text). Kunz\cite{Kun82} performed band structure calculations for the solid
and CI represents a Configuration-Interaction cluster calculation\cite{Ber96}.
All energies are given in eV.

$\;$

{\bf Table II}
Band gaps of seven oxides
calculated from several theoretical methods, compared to experimental
values.
$\Delta \epsilon_{HF}$ is a ($LUMO-HOMO$) difference of
one-particle orbital energy eigenvalues. $\Delta$SCF refers to a difference
between the energies of the crystal in appropriate electronic states (see
text).
Also included are the {\em ab initio} band structure calculations of Pandey
{\em et al}\cite{Pan91} and the {\em ab initio} valence bond cluster model
results of Lorda {\em et al}\cite{Lor93}.
All energies are given in eV.

\onecolumn[\hsize\textwidth\columnwidth\hsize\csname
@onecolumnfalse\endcsname

\begin {table}
\begin {center}
\begin {tabular} {|c|c|c|c|c|c|c|c|c|c|}
\hline
& &
\multicolumn {2}{c|}{$\Delta \epsilon_{HF}$(LUMO-HOMO)} &
\multicolumn {3}{c|}{$\Delta$SCF} &
\multicolumn {2}{c|}{$Others$} & 
\multicolumn {1}{c|}{} \\ 
\hline
 & $Crystal$ & This work & ref.[52] & HF & CHF & ref.[52] & Kunz & CI[52] & Exp. \\ 
\hline
& $LiF$ & 14.9 & 16.1 & 12.3 & 14.1 & 12.2 & 14.0 & - & 14.2  \\
& $NaF$ & 13.6 & 14.7 & 10.1 & 12.6 & 11.6 & 12.0 & 11.9 & 11.5 \\ 
& $KF$ & 12.4 & 13.3 & 10.5 & 11.8 & 10.6 & 10.9 & 11.6 & 10.8  \\ 
& $RbF$ & 12.0 & 12.7 & 9.3 & 10.7 & 10.1 & 11.0 & 11.3 & 10.3  \\ 
& $CsF$ & 11.6 & - & 8.7 & 10.5 & - & - & - &  9.9  \\ 
& $LiCl$ & 11.2 & 13.0 & 10.15 & 9.15 & 10.4 & 9.7 & - &  9.4 \\
& $NaCl$ & 10.8 & 12.2 & 9.6 & 8.5 & 9.8 & 10.0 & 8.9 &  9.0  \\
& $KCl$ & 9.9 & 11.2 & 8.7 & 7.6 & 9.2 & 10.0 & 8.8 &  8.7  \\
& $RbCl$ & 9.6 & 10.8 & 8.2 & 7.4 & 9.0 & 10.3 & 8.7 &  8.5   \\
& $CsCl$ & 9.0 & - & 7.8 & 6.7 & - & - & - &  8.3  \\
& $LiBr$ & 10.9 & 12.0 & 10.5 & 8.2 & 9.6 & 8.4 & - &  7.6  \\
& $NaBr$ & 10.5 & 11.4 & 9.8 & 7.0 & 9.0 & 10.0 & 7.9 &  7.1  \\
& $KBr$ & 9.7 & 10.5 & 9.1 & 6.5 & 8.6 & 8.7 & 7.9 &  7.4  \\
& $RbBr$ & 9.4 & 10.2 & 8.8 & 6.2 & 8.2 & 7.1 & 7.8 &  7.2   \\
& $CsBr$ & 8.9 & - & 8.3 & 5.4 & - & - & - &  7.3  \\
& $LiI$ & 10.0 & 11.0 & 9.5 & 7.4 & 8.8 & - & - &  6.4  \\
& $NaI$ & 9.7 & 10.5 & 9.0 & 6.9 & 8.4 & - & 6.9 &  6.0  \\
& $KI$ & 8.9 & 9.7 & 8.4 & 6.3 & 7.9 & - & 7.0 &  6.1  \\
& $RbI$ & 8.7 & 9.4 & 8.0 & 5.9 & 7.7 & - & 6.9 &  5.8   \\
& $CsI$ & 8.3 & - & 7.7 & 5.6 & - & - & - &  6.2 

\end {tabular}
\end {center}

\caption {}

\end {table}

\begin {table}
\begin {center}
\begin {tabular} {|c|c|c|c|c|c|c|c|}
\hline
& $Crystal$ & $\Delta \epsilon_{HF}$ & $\Delta$SCF(HF) & $\Delta$SCF(CHF) & Pandey & Lorda & Exp. \\
\hline
& $BeO$ &  10.29 & 8.49 & 10.84 & -  &  -   & 10.6 $^a$ \\
& $MgO$ &  7.91 & 5.80 & 7.99 & 8.21 & 10.9 & 7.69 $^b$ \\ 
& $CaO$ &  7.14 & 5.08 & 7.01 & 7.73 & 8.0  & 7.0 $^b$ \\ 
& $SrO$ &  6.15 & 3.86 & 5.77 & 7.1  & 6.2  & 5.7 $^b$ \\ 
& $BaO$ &  4.92 & 2.76 & 4.62 &  -   & 4.8  & 4.8 $^b$ \\ 
& $SiO_2$ & 9.17 & 6.76 & 8.67 &  -   &  -   & $\sim$ 9 $^c$ \\
& $Al_2O_3$ & 8.96 & 6.93 & 9.23 &  -   &  -   & $>$ 9 $^d$ \\
\end {tabular}
\end {center}

$^a$ Ref.\cite{Web86}

$^b$ Ref.\cite{Str73}

$^c$ Ref.\cite{Ara68}

$^d$ Ref.\cite{Him86}

\caption {}

\end {table}


\begin{thebibliography}{99}

\bibitem{Bla95} G. BLASSE and B. C. GRABMAIER, Luminescent materials, Springer Verlag, Berlin (1995).

%
\bibitem{Ber95} M. BERRONDO and J. F. RIVAS-SILVA, {\em Int. J. Quantum Chem.
Quantum Chem. Symp.} {\bf 29},
                  253 (1995).

\bibitem{Huz71} S. HUZINAGA and A. P. CANTU, {\em J. Chem. Phys.} {\bf 55},
                 5543 (1971).

\bibitem{Huz73} S. HUZINAGA, D. MCWILLIAMS, and A. A. CANTU, {\em Adv. Quantum Chem.}
                 183 (1973).

\bibitem{Lua87} V. LUA\~NA. {\em Thesis Disertation}, Universidad de 
                Oviedo (1987).

\bibitem{Lua90} V. LUA\~NA and L. PUEYO, {\em Phys. Rev. B} {\bf 41},
                  3800 (1990).

\bibitem{Ayu93} A. AYUELA, J. M. L\'OPEZ, J. A. ALONSO, and V. LUA\~NA,
                {\em Z. Phys. D} {\bf 26}, S213 (1993); ({\em ibid.}) Physica B {\bf 212},
                329 (1995); ({\em ibid.}) {\em Canad. J. Phys.} (in press).

\bibitem{Agu97} A. AGUADO, A. AYUELA, J. M. L\'OPEZ, and J. A. ALONSO,
                {\em J. Phys. Chem.} {\bf 101B}, 5944 (1997);
                {\em Phys. Rev. B}, {\bf 56}, 15353 (1997);
                E. DE LA PUENTE, A. AGUADO, A. AYUELA, and J. M. L\'OPEZ,
                {\em Phys. Rev. B} {\bf 56}, 7607 (1997).

\bibitem{Cla68} T. D. CLARK and K. L. KLIEWER, {\em Phys. Lett.} {\bf 27A},
                  167 (1968).

\bibitem{Kun68a} A. B. KUNZ, {\em Phys. Rev.} {\bf 175},
                  1147 (1968).

\bibitem{Fon68} C. Y. FONG and M. L. COHEN, {\em Phys. Rev. Lett.} {\bf 21},
                  22 (1968).

\bibitem{Kun68b} A. B. KUNZ, {\em Phys. Stat. Sol. (b)} {\bf 29},
                  115 (1968).

\bibitem{Fow69} W. B. FOWLER and A. B. KUNZ, {\em Phys. Rev.} {\bf 186},
                  956 (1969).

\bibitem{Pag70} L. J. PAGE and E. H. HYGH, {\em Phys. Rev. B} {\bf 1},
                  3472 (1970).

\bibitem{Kun70} A. B. KUNZ, {\em J. Phys. C: Solid St. Phys.} {\bf 4},
                  1542 (1970).

\bibitem{Ove71} H. OVERHOF, {\em Phys. Stat. Sol. (b)} {\bf 43},
                  575 (1971).

\bibitem{Lip71} N. O. LIPARI and A. B. KUNZ, {\em Phys. Rev. B} {\bf 3},
                  491 (1971).

\bibitem{Per72} F. PERROT, {\em Phys. Stat. Sol. (b)} {\bf 52},
                  163 (1972).

\bibitem{Poo75a} R. T. POOLE, J. LIESEGANG, R. C. G. LECKEY, and J. G. JENKINS,
                 {\em Phys. Rev. B} {\bf 11},
                   5190 (1975).

\bibitem{Jou75} C. JOUANIN, J. P. ALBERT, and C. GOUT, {\em Il Nuovo Cimento}
                {\bf 28},
                  483 (1975).

\bibitem{Kun82} A. B. KUNZ, {\em Phys. Rev. B}
                {\bf 26}, 2056 (1982).

\bibitem{Hea83} R. A. HEATON, J. C. HARRISON, and C. C. LIN, {\em Phys. Rev. B} 
                {\bf 28},
                  5992 (1983).

\bibitem{Pan91} R. PANDEY, J. E. JAFFE, and A. B. KUNZ, {\em Phys. Rev. B} 
                {\bf 43}, 9228 (1991).

\bibitem{Lyk56} P. G. LYKOS and R. G. PARR, {\em J. Chem. Phys.} {\bf 24},
                  1166 (1956).

\bibitem{Par56} R. G. PARR, R. O. ELLISON, and P. G. LYKOS, {\em J. Chem. Phys.}
                {\bf 24}, 1106 (1956).

\bibitem{Huz87} S. HUZINAGA, L. SEIJO, Z. BARANDIAR\'AN, and M. KLOBUKOWSKI,
                {\em J. Chem. Phys.} {\bf 86}, 2132 (1987).

\bibitem{Sil77} H. J. SILVERSTONE and R. K. MOATS, {\em Phys. Rev. A}
                {\bf 16}, 1731 (1977). 

\bibitem{Map91} A. MART\'IN PEND\'AS and E. FRANCISCO, {\em Phys. Rev. A}
                {\bf 43}, 3384 (1991). 

\bibitem{Bar83} Z. BARANDIAR\'AN and L. PUEYO, {\em J. Chem. Phys.} {\bf 79},
                  1926 (1983).

\bibitem{Kun88} A. B. KUNZ and J. M. VAIL, {\em Phys. Rev. B} {\bf 38},
                  1058 (1988).

\bibitem{McW94} R. MCWEENY,
                {\em Methods of molecular quantum mechanics}, 
                Academic Press. London (1994).

\bibitem{Cle65} E. CLEMENTI, {\em IBM J. Res. Develop.},
                {\bf 9}, 2 (1965).

\bibitem{Cha89} S. J. CHAKRAVORTY and E. CLEMENTI, {\em Phys. Rev. A},
                {\bf 39}, 2290 (1989).

\bibitem{Roo63} C. C. J. ROOTHAAN and P. S. BAGUS,
                {\em Methods Comput. Phys.},
                {\bf 2}, 47 (1963).

\bibitem{Pue92} L. PUEYO, V. LUA\~NA, M. FL\'OREZ, and E. FRANCISCO 
                in {\em Structure, Interactions and Reactivity},
                edited by S. FRAGA (Elsevier, Amsterdam, 1992), vol. B, p. 504.

\bibitem{Lua92a} V. LUA\~NA, M. FL\'OREZ, E. FRANCISCO, A. MART\'IN
                 PEND\'AS, J. M. RECIO, M. BERMEJO, and L. PUEYO, in {\em Cluster
                 Models for Surface and Bulk Phenomena}, edited by G. PACCHIONI,
                 P. S. BAGUS and F. PARMIGIANI (Plenum, New York, 1992), p. 605.

\bibitem{Cle74} E. CLEMENTI and C. ROETTI, {\em At. Data Nucl. Data Tables},
                {\bf 14}, 3 (1974); {\bf 14}, 177 (1974). 

\bibitem{Mcl81} A. D. MCLEAN and R. S. MCLEAN, {\em At. Data Nucl. Data Tables},
                {\bf 26}, 197 (1981). 

\bibitem{Bag94} P. S. BAGUS, F. ILLAS, and C. SOUSA,
                {\em J. Chem. Phys.} {\bf 100}, 2943 (1994).

\bibitem{Poo75} R. T. POOLE, J. G. JENKINS, R. C. G. LECKEY, and J. LIESEGANG
                 {\em Phys. Rev. B} {\bf 11}, 5179 (1975).

\bibitem{Bun93} C. F. BUNGE and J. A. BARRIENTOS, {\em At. Data Nucl. Data
                Tables} {\bf 53}, 113 (1993). 

\bibitem{Raf73} R. C. RAFFENETTI and K. RUEDENBERG. {\em Even-Tempered Representations of Atomic Self-Consistent-Field Wave Functions} (Ames Laboratory Report, 1973).

\bibitem{Cas93} J. CASANOVAS, C. SOUSA, J. RUBIO, and F. ILLAS,
                {\em J. Comput. Chem.} {\bf 14}, 680 (1993).

\bibitem{Hay85} W. HAYES and A. M. STONEHAM, {\em Defects and Defect Processes in 
                Non-Metallic Solids},
                Wiley, New York (1985).

\bibitem{Web86} {\em Handbook of Laser Science and Technology},
                edited by M. J. WEBER (CRC, Cleveland, 1986) Vol. III.

\bibitem{Str73} W. H. STREHLOW and E. L. COOK,
                {\em J. Phys. Chem., Ref. Data} {\bf 2}, 163 (1973).

\bibitem{Ara68} E. T. ARAKAWA and M. W. WILLIAMS,
                {\em J. Phys. Chem. Solids} {\bf 29}, 735 (1968).

\bibitem{Him86} F. J. HIMPSEL,
                {\em Surf. Sci.} {\bf 168}, 764 (1986);
                F. J. GRUNTHONER and P. J. GRUNTHONER,
                {\em Mater. Sci. Rep.} {\bf 1}, 65 (1986).

\bibitem{Lor93} A. LORDA, F. ILLAS, J. RUBIO, and J. B. TORRANCE,
                {\em Phys. Rev. B} {\bf 47}, 6207 (1993).

\bibitem{Wyc63} R. W. G. WYCKOFF,
                {\em Crystal Structures, Volume I, New York: Wiley-Interscience}
                (1963).

\bibitem{Kit76} C. KITTEL, {\em Introduction to Solid State Physics},
                 Wiley, New York (1976).

\bibitem{Ber96} M. BERRONDO and J. F. RIVAS-SILVA.
                {\em Int. J. Quantum Chem.}, {\bf 57}, 1115 (1996).

\bibitem{God88} R. W. GODBY, M. SCHL\"UTER, and L. J. SHAM,
                {\em Phys. Rev. B} {\bf 37}, 10159 (1988);
                L. J. SHAM and M. SCHL\"UTER,
                {\em Phys. Rev. B} {\bf 32}, 3883 (1985).

\bibitem{Map92} A. MART\'IN PEND\'AS, E. FRANCISCO, V. LUA\~NA, and L.PUEYO,
                {\em J. Phys. Chem.} {\bf 96}, 2301 (1992).

\bibitem{Pan74} S. T. PANTELIDES, D. J. MICKISH, and A. B. KUNZ,
                {\em Phys. Rev. B} {\bf 10}, 2602 (1974).

\bibitem{Mar83} Theory of the Inhomogeneus Electron Gas.
                Editors S. LUNDQVIST and N. H. MARCH, Plenum, New York (1983)

\bibitem{Hyb86} M. S. HYBERTSEN and S. G. LOUIE,
                {\em Phys. Rev. B} {\bf 34}, 5390 (1986).

\bibitem{Hed65} L. HEDIN,
                {\em Phys. Rev.} {\bf 139}, A796 (1965);
                L. HEDIN and S. LUNDQVIST,
                {\em Solid State Phys.} {\bf 23}, 1 (1969).

\end{thebibliography}
\end{document}